# Pump-locked microcavity Brillouin laser

Yuqin Mao,[1,2] Chaoze Zhang,[1,2] Ligang Huang,[1,*] Lei Gao,[1] Yujia Li,[1] Leilei Shi,[1] Guolu Yin,[1] Chaoyang Gong,[1] and Tao Zhu[1,3]

[1]Key Laboratory of Optoelectronic Technology & Systems (Ministry of Education), Chongqing University, Chongqing 400044, China

[2]These authors contributed equally

[3]zhutao@cqu.edu.cn

*Corresponding author: lghuang@cqu.edu.cn

Abstract – Microcavity-based microlasers are the kernel light sources for integrating photonics and optoelectronics. The traditional pump light frequency locking mainly utilizes a complex system with optoelectronic feedback, which requires a high-cost narrow-linewidth pump laser and limits the application of microlasers in integrated optoelectronic systems. We propose to utilize Rayleigh scattering of microcavities to lock the frequency of the pump laser to the resonant frequency of the laser microcavity with an all-optical method. While compressing the linewidth of the pump laser, it can greatly improve the long-term stability of the optically pumped microcavity laser. In the experiment, the linewidth of the semiconductor pump laser is compressed from the MHz level to the kHz level. The microcavity Brillouin laser achieves an ultra-narrow intrinsic linewidth of 100 Hz, with an ultra-low frequency noise of 35 $Hz^2$/Hz. The constructed microlaser obtains a locking time up to 1 hour, which does not require any temperature control or vibration isolation of the laser system. This work is the first demonstration to achieve an optically pump-locked microcavity Brillouin laser, which provides a stable and reliable low-cost experimental platform for ultra-narrow linewidth lasers, precision laser sensors, microwave-photonic signal synthesizer, and optomechanical systems.

Index Terms: whispering gallery mode, stimulated Brillouin scattering, microcavity, narrow-linewidth laser, microlaser.

1. Introduction

Optical microcavities, distinguished for its capability to enhance light-matter interactions within extremely small mode volumes, have a wide range of applications in various fields, including cavity quantum electrodynamics (CQED) [1-7], optomechanical systems [8-13], microlaser sources [14-24], optical signal processing [25-29] and precision sensing [30-35]. In particular, the microcavity-based microlaser can exhibit much narrower linewidth than the passive microcavity, which can obtain better sensitivity such as in biochemical sensing [36-37]. Additionally, microcavity-based microlasers serve as important light sources in integrated photonics and optoelectronics [38-40]. Optical pumping is a very important method that can be used for stimulated Brillouin scattering (SBS), stimulated Raman scattering (SRS), parametric down-conversion, rare-earth-doped microlasers, which are required to lock the main cavity laser with the pump laser in order to improve the pumping efficiency [41-45]. Among the microlasers based on different gain mechanisms, SBS microlasers can produce ultra-narrow-linewidth photons and long-life-time microwave-band phonons simultaneously, which gain unique applications in integrated microwave photonics (MWP), optomechanics, gyroscopes and precision sensors [46-51]. As the gain bandwidth of SBS is extremely narrow (in MHz level and below), locking the pump laser becomes extremely difficult and important. One approach to lock the pump laser frequency is the Pound–Drever–Hall (PDH) servo loop of optoelectronic feedback, which requires high-cost narrow-linewidth tunable pump lasers, complex servo feedback systems and bulk laser packaging structures. Another approach is to use thermal self-locking of the microcavity, which will lead the resonant frequency of the microcavity to follow the fluctuating pump laser frequency. Thermal self-locking of the microcavity still requires high stability of the pump laser frequency. Otherwise, the microcavity will lose lock due to the instantaneous frequency jitter of the pump laser. Even in the locked state, the resonant frequency of the microcavity will still drift with the jitter of the pump laser frequency, which is not conducive to the stable operation of the SBS microlaser and induces the laser linewidth broadening. It is worth noting that the microcavity can serve not only as an independent resonant cavity but also as a resonant feedback device for an external-cavity laser [52]. Long-term locking of the external-cavity laser frequency can be achieved by injecting the backward Rayleigh scattering from the microcavity, and the locked frequency mainly depends on the resonant frequency of the microcavity. Furthermore, Rayleigh scattering has also been proved to be an effective way to deeply compress laser linewidth [53-61]. Therefore, utilizing all-optical feedback based on Rayleigh scattering of microcavity is promising to achieve the frequency locking of the pump laser and form more stable microlaser systems, including microcavity Brillouin lasers.

In this work, we propose an all-optically pump-locked microcavity Brillouin laser, as a typical microlaser to demonstrate the laser stability enhancement ability from all-optical pump locking. Besides compressing the linewidth of the pump laser, the Rayleigh scattering from the microcavity can greatly improve the long-term stability of the microcavity Brillouin laser. In the experiment, the linewidth of the pump laser can be compressed from MHz level to kHz level based on the feedback of Rayleigh scattering from the microlaser cavity, with the white frequency noise floor reduced from 7160 $Hz^2$/Hz to 7 $Hz^2$/Hz, which is compressed by 3 orders. On the condition of pump locking, the microlaser operates as a stable Brillouin laser with intrinsic linewidth as narrow as 100 Hz and frequency noise as low as 35 $Hz^2$/Hz. Benefiting from the ultra-high stability of the all-optical pump locking, the wavelength the Brillouin microlaser can be step-by-step tuned from 1549.3 nm to 1550 nm, by tuning the wavelength of pump laser to stimulate the SBS at different resonant modes of the microcavity. The wavelength of the Brillouin microlaser can be further continuously tuned within 20 pm, by tuning the pump power which can finely tuning the resonant wavelength of the microcavity based on the thermal-optic effect. The pump-locked state of the constructed Brillouin microlaser can maintain at least 1 hour under normal laboratory conditions. It is worth noting that the whole experiment was carried out without thermostatic or vibration isolation, which indicates that the pump-locked microcavity Brillouin laser has a great long-time stability. This work is the first demonstration to achieve an optically pump-locked microcavity Brillouin laser. The proposed pump-lock method greatly improves the stability of optically pumped microcavity lasers and does not require expensive narrow-linewidth tunable lasers, which largely reduces the complexity and cost of the practical microlaser system. The constructed pump-locked microcavity Brillouin laser provides a stable and reliable low-cost experimental platform for ultra-narrow linewidth lasers, precision laser sensors, microwave-photonic signal synthesizer, and optomechanical systems.

2. Principle and configuration

The generation process of SBS is that an acoustic phonon and a frequency-shifted Stokes photon are released after the annihilation of a pump photon during the interaction of the pump light wave with the gain medium [9], as shown in Fig. 1(a). Thanks to the high Q and the very small mode volume of a WGM microcavity such as the microsphere resonator, the threshold for generating SBS in the microcavity is significantly reduced. When the pump laser that reaches the power of the SBS threshold condition and satisfies the microcavity resonant frequency is coupled into the microcavity through the tapered fiber, Stokes light can be produced. At this point, if a certain WGM resonance frequency happens to exist within the Brillouin gain bandwidth, the Stokes light at this frequency can be nonlinearly amplified to produce SBS, as shown in Fig. 1(b). The propagation path of the SBS generated in the microsphere cavity is total reflection along the equatorial direction on the inner surface of the microsphere cavity, and the propagation path of the light wave inside the microsphere cavity is shown in Fig. 1(c). Using the classical electromagnetic theory, one can obtain the field distribution of the WGM by solving the set of Maxwell's equations for the specific boundary shape of the microsphere resonator. By varying the radial mode number (*n*), the polar mode number (*l*) and the azimuthal mode number (*m*) of the microsphere, we can obtain different mode field distributions inside the microcavity. Figure 1(d) gives the mode field distribution of the microsphere cavity in the cross-section at the coupling point along the propagation direction of the light wave at different *n, l* and *m*. Considering a microsphere resonator with a large diameter (>100 μm), when a certain ellipticity exists in it, the WGM resonant wavelengths with different *m* will be separated from each other [62], and thus a large number of resonant peaks can be formed, which will give a greater possibility for the frequency-matching condition of SBS. For the excitation of the WGM in the microsphere, we use a tapered fiber with a diameter range of 1-2 μm in the beam waist region to couple with a silica microsphere cavity with a diameter of about 200 μm. With fast scanning of the microsphere cavity by a tunable laser with low power, the multi-peak transmission spectrum of the microsphere cavity is obtained as shown in Fig. 1(e), which shows that the microsphere cavity has a large number of resonance peaks. Figure 1(f) is one of the resonant peaks in the above multi-peak transmission spectrum, which has a full-width at half maximum (FWHM) of 1 MHz, corresponding to a Q factor of $2\times10^8$ for the microsphere cavity.

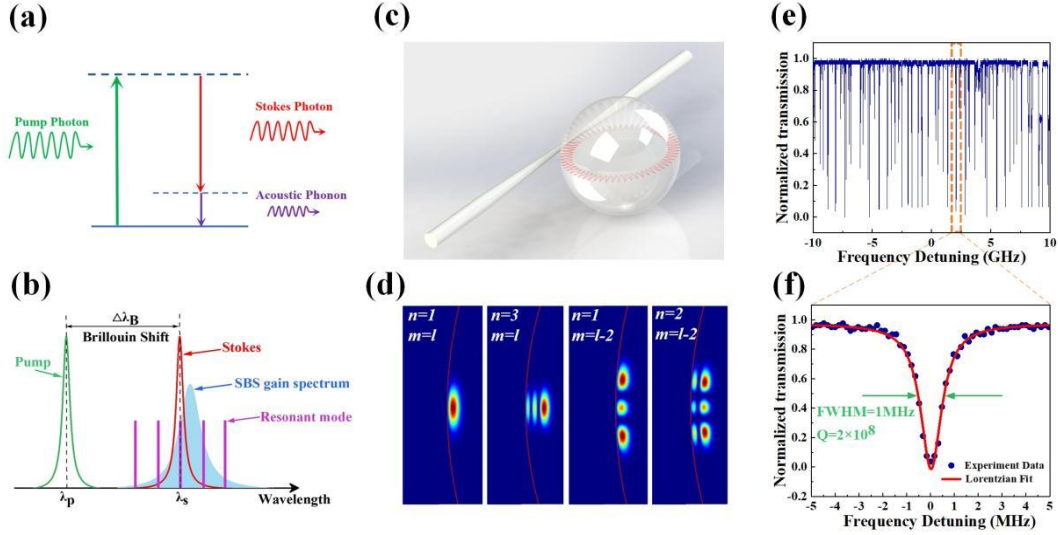

Fig. 1. The basic physical process of generating SBS and the performance test of the microsphere cavity. (a) Energy level transition of particles during Stokes photon generation. (b) Brillouin gain process in the microsphere cavity. (c) Total reflection of light waves entering the microsphere cavity on its inner surface. (d) Distribution of different mode fields in the cross-section of the propagation direction at the coupling site. (e) Multi-peak transmission spectrum of the microsphere cavity and (f) one of the resonance peaks, indicating that the microsphere cavity has a Q factor of 2×10$^8$.

The schematic of the proposed pump-locked SBS microlaser based on a WGM microsphere cavity is shown in Fig. 2(a). A distributed feedback (DFB) semiconductor laser chip without an isolator is used as the optical pump. The DFB pump laser is focused through a convex lens into a 90/10 optical coupler (OC), which is connected to a tapered fiber. The DFB pump laser couples with the microsphere cavity through the tapered fiber. Figure 2(b) shows the coupling system consisting of a microsphere cavity and a tapered fiber, with the propagation of pump, Rayleigh scattering (RS), and SBS. Due to the roughness of the surface of the microsphere cavity and the inhomogeneity of its internal material, backward RS is formed, and part of the RS returns to the DFB laser, which constitutes a self-injection locking mechanism. For an external cavity laser diode, the linewidth Δv$_p$ is [63]

$$\Delta v_p = \frac{\pi h v^3 n_{sp}}{PQQ_E}(1+\alpha_H^2), \qquad (1)$$

where $h$ is Planck's constant, $v$ is the laser frequency, $n_{sp}$ is the population inversion factor, $P$ is the total emitted output power, $Q$ and $Q_E$ are the loaded and external quality factors of the laser cold cavity, and $\alpha_H$ is the linewidth enhancement factor. According to Eq. (1), a smaller linewidth can be realized when an external cavity with a higher Q factor. The WGM microsphere cavity with a $Q$ factor as high as 2×10$^8$ is used in the proposed configuration, and the linewidth of the DFB pump laser can be narrowed due to the resonant feedback from this microcavity. In the experiment, the output power of the DFB is around 20 mW, and the high power pump increases the temperature inside the microsphere cavity, which causes the resonant wavelength of the microsphere cavity to collectively redshift due to the thermal effect of the microsphere cavity itself. Simultaneously, the wavelength of the DFB pump laser is tuned to the long wavelength direction to ensure that both the pump and Stokes light can catch up with the redshifted resonant wavelengths of the microsphere cavity. When the Stokes light catches up with a specific resonant mode of the microsphere cavity, the SBS laser can be generated and forms a thermal self-locked state. There are two main fundamental noise sources, i.e. the pump noise and Schawlow-Towns (ST) noise, both determining the SBS laser linewidth as [22,64,65]:

$$\Delta v = \frac{\Delta v_p}{(1+\Gamma_A/\gamma_B)^2} + \frac{\hbar\omega^3}{4\pi P_{out} Q_{tot} Q_{ext}}(n_T + N_T + 1). \qquad (2)$$

In the first term, $\Delta v_p$ is the linewidth of the pump laser, and $\Gamma_A$ and $\gamma_B$ are the damping rates of the acoustic mode and the optical Stokes mode respectively. The second term in Eq. (2) refers to the quantum noise, or ST-like noise for it is inversely proportional to the output power $P_{out}$. $Q_{tot}$ and $Q_{ext}$ are the total and the external $Q$ factors, respectively. $n_T$ and $N_T$ refer to the numbers of thermal quanta in the mechanical field and the optical field. From Eq. (1), the linewidth of the DFB pump laser is effectively narrowed when using a microsphere cavity with a high Q factor, i.e. $\Delta v_p$ in the first term of Eq. (2) is reduced, while the second term of Eq. (2) can also be reduced by increasing the $Q$ factor. Therefore, self-injection locking of the pump laser to the high-$Q$ microsphere cavity by the backward RS can effectively compress the linewidth of the SBS microlaser. In the meantime, the DFB pump and the microsphere cavity form a mutual optical locking, which can largely enhance the transient frequency stability of the DFB pump laser, and no additional optoelectronic feedback technique is required in the whole locking process. Even when the temperature in the environment drifts, the self-injection locking effect can still lock the pump laser frequency to the frequency-drifting resonant mode of the microsphere, which largely enhances the locking stability of pump laser to the microresonator.

3. Experimental results and discussion

In the experiment, the delayed-self-heterodyne interferometer (DSHI) measurement system is used to test the laser performance, and Figs. 2(c)-2(e) show comparisons of the transient frequency, linewidth, and frequency noise of the DFB pump laser before and after self-injection locking. The demodulated frequency jitter range of the DFB pump after locking is reduced from ±~1.5 MHz to ±~0.08 MHz within a time window of 10 ms. In the meantime, the linewidth is narrowed from 54.7 kHz to 920 Hz, and the frequency noise floor is reduced from 7160 Hz$^2$/Hz to 7 Hz$^2$/Hz. After frequency stabilization and linewidth compression of the DFB pump laser based on the self-injection locking to the microsphere, the SBS laser of the microresonator can be then stimulated more efficiently. To observe the SBS laser performance, the reflective light from the tapered-fiber- coupled microsphere is poured to a tunable bandpass filter to select the Stokes laser and remove the backward RS light from the microsphere with the pump laser frequency. Figure 2(f) shows the output spectra of the constructed microlaser. The black curve denotes the output spectrum of the DFB pump before locking to the microsphere. The blue curve denotes the output spectrum of the unfiltered reflective light from the microsphere after locking the pump, which shows obvious two-wavelength feature with interval of 0.09 nm, meaning that the reflective light contains Stokes laser and the backward RS light simultaneously. The red curve denotes the output spectrum of the reflective light after filtering, showing that the Stokes laser is accurately selected, which can be further poured to the linewidth and noise measurement system. The measured linewidth and frequency noise of the Brillouin laser are shown in Figs. 2(g) and 2(h), respectively. The linewidth can be calculated to be 14 kHz/20=700 Hz based on the 20 dB linewidth, which reaches sub-kHz level, while the frequency noise floor is 35 Hz$^2$/Hz. Based on the β-separation line as denoted by the blue line in Fig. 2(h), the integrated linewidth can be estimated in kHz level [66].

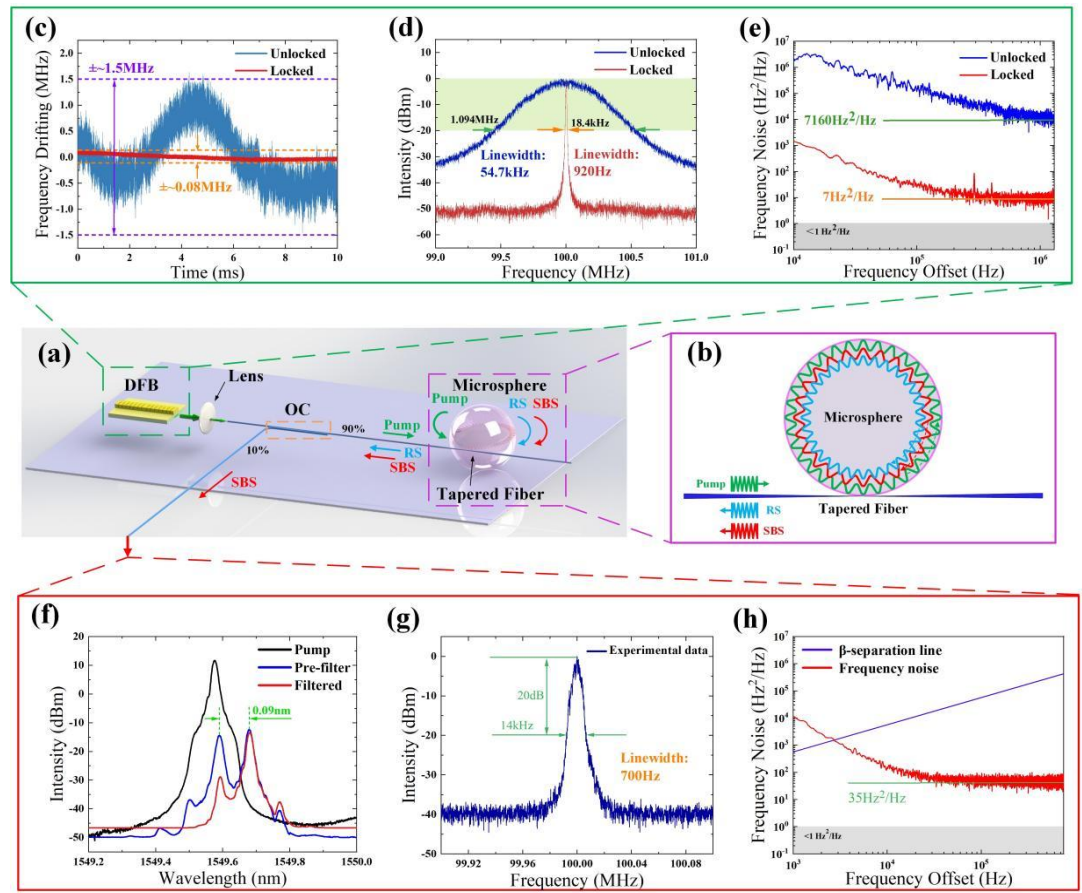

Fig. 2. (a) Schematic of the pump-locked Brillouin microlaser. (b) The stimulated light distribution of tapered-fiber-coupled microsphere, including the pump (green), RS (blue) and SBS (red). (c) The transient frequency, (d) linewidth measurement and (e) frequency noise of the DFB pump laser before and after locking. (f) Spectra of the pump laser and Brillouin laser. (g) Linewidth measurement and (h) frequency noise of the Brillouin microlaser.

The wavelength tunability of the Brillouin microlaser is tested by tuning the wavelength of the DFB pump laser. The temperature of DFB laser chip can be controlled by the thermoelectric cooler (TEC) and the thermistor as temperature sensor, which can be utilized to precisely control the resonant wavelength of the gratings in DFB chip. In the wavelength tuning process, the pump current of the DFB laser is set to be 200 mA, and the output power of the laser chip maintains about 20 mW. Figure 3(a) shows the spectra of the filtered Stokes laser during the wavelength tuning. The wavelength of the Brillouin laser can be tuned from 1549.3484 nm to 1549.9771 nm. It is worth noting that the Brillouin microlaser only can be tuned to several discrete wavelengths, including 1549.3484 nm, 1549.3705 nm, 1549.4706 nm, 1549.6267 nm, 1549.6728 nm, 1549.7969 nm and 1549.9771 nm, when maintaining the pump current to be 200 mA. The strict

step-tuning feature of the Brillouin microlaser is obtained, because the wavelength of the pump laser has to be strictly locked within the discrete resonant peaks of the microsphere. In fact, the output wavelength of the locked DFB pump laser is determined by the resonance wavelength of the microsphere. When the pump power maintains constant, the inner temperature of the microsphere will maintain stable, and the resonant wavelength of the microsphere will maintain stable, accordingly. Therefore, when tuning the temperature and resonance wavelength of the DFB gratings, the output wavelengths of the pump DFB laser and the Brillouin microlaser are both locked to the stable and discrete resonance wavelengths of the microsphere. Throughout the wavelength tuning process, the linewidth and frequency noise of the Brillouin microlaser are also measured, as shown in Figs. 3(b) and 3(c), respectively. Figure 3(d) summarizes the variation of the linewidth and frequency noise floor. The linewidth of the SBS laser fluctuates between 600 Hz and 830 Hz, and the frequency noise floor fluctuates between 27 $Hz^2/Hz$ and 90 $Hz^2/Hz$. The fluctuation mainly originates from the different $Q$ factors, when locking the Stokes laser to different WGMs with different resonance wavelengths in the microsphere.

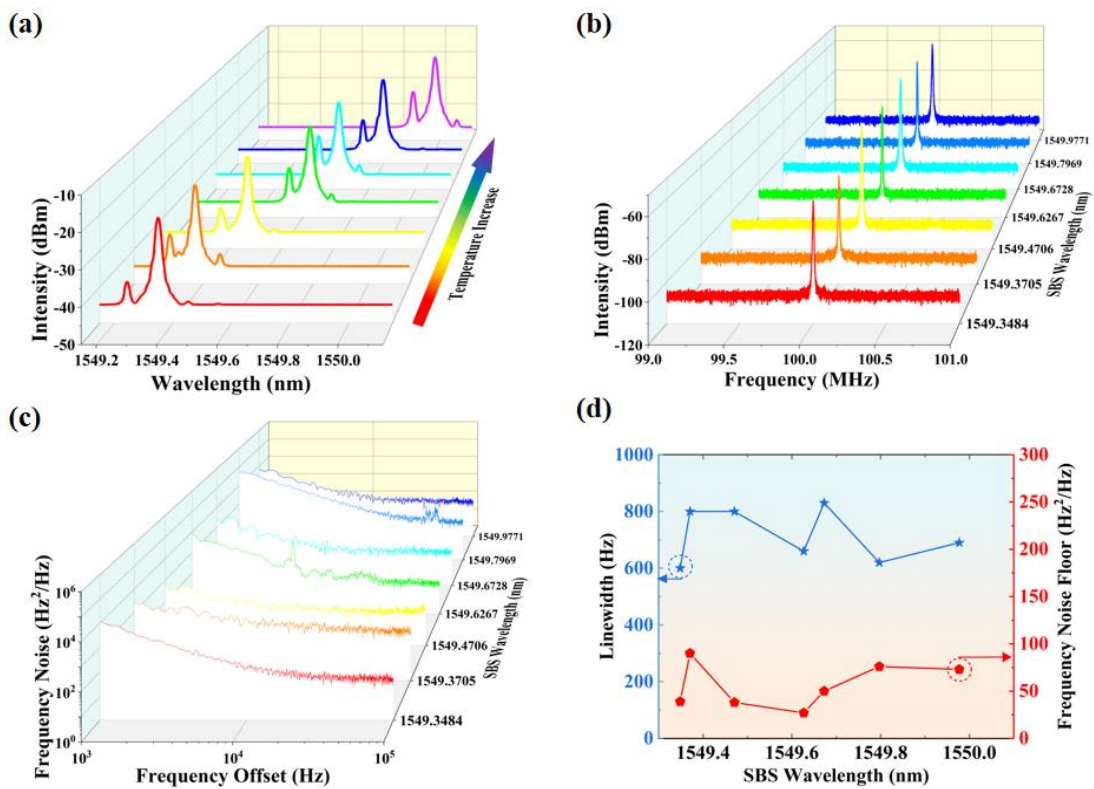

Fig. 3. Wavelength tunability of the Brillouin microlaser. (a) Tuned Stokes laser spectra. (b) Linewidth measurement of tuned wavelengths. (c) Frequency noise of tuned wavelengths. (d) Measured linewidth and frequency noise floor of tuned wavelengths.

The power tunability of the SBS microlaser is tested by tuning the output power of the DFB pump laser. The output power is increased by increasing the pump current of the DFB pump laser. Figure 4(a) shows the spectra of the filtered SBS microlaser when tuning pump current from 214.7 mA to 228.1 mA. In the meantime, the linewidth is measured by DSHI method as shown in Fig. 4(b). The measured peak intensity of the spectrum and linewidths at different pump currents are shown in Fig. 4(c). The output intensity of the SBS microlaser is increased from -28 dBm to

-13 dBm, with the pump current tuned from 214.7 mA to 222.3 mA, and maintains about -13 dBm when further increasing the pump current. At the same time, the linewidth of the SBS microlaser decreases from the initial 45 kHz to 700 Hz, with the pump current tuned from 214.7 mA to 222.3 mA, and maintains about 700 Hz when further increasing the pump current. The SBS microlaser is in an unstable state when tuning the pump current between 214.7 mA and 221.2 mA, because the intensity in the microsphere is still unstable near the threshold condition of the SBS laser. It is worth noting that the wavelengths of the DFB pump laser and the SBS microlaser are both slightly drifting when increasing the pump current, as shown in Fig. 4 (d). The wavelength of the DFB pump laser and the SBS microlaser can be continuously tuned from 1549.1512 nm to 1549.2152 nm, and from 1549.2402 nm to 1549.3023 nm, respectively, when the pump current is tuned from 214.7 mA to 228.1 mA. The wavelength interval between the pump and Stokes light maintains about 0.09 nm (~11 GHz) in the fine tuning process, which is in accordance with the theory of the Brillouin frequency shift in silica [9]. The wavelength tuning when increasing the pump current is induced by the thermal-optical effect, because the increased pump power will induce the temperature increase of the microsphere.

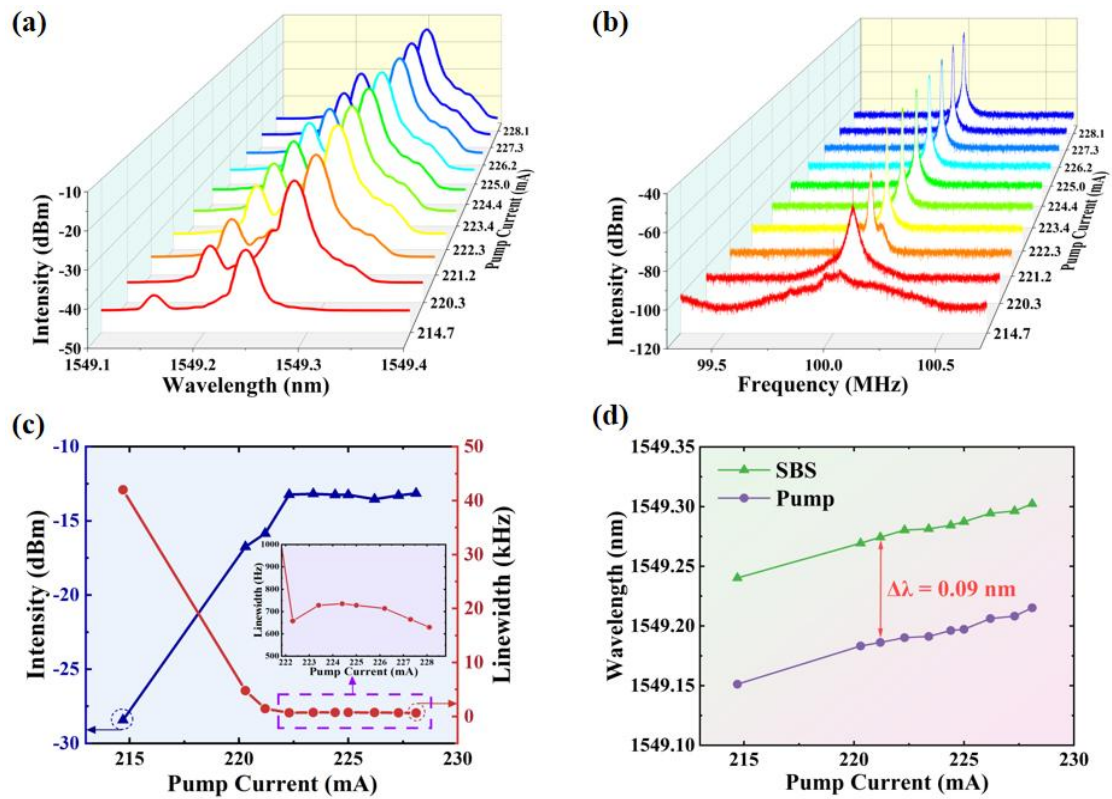

Fig. 4. Power tunability of the Brillouin microlaser. (a) Tuned spectra. (b) Linewidth measurement under different pump currents. (c) Relationship between the spectrum peak intensity, linewidth and the pump current. (d) Fine tuning of the wavelength of the DFB pump laser and the SBS microlaser.

The stability of the laser wavelength, output power and linewidth are tested at 5-minute interval within 1 hour. Figure 5 (a) and 5 (b) shows the output optical spectra and DSHI spectra of the Brillouin microlaser, respectively. The laser wavelength and intensity stability can be then obtained from the laser optical spectrum, as shown in Fig. 5(c). The fluctuations of the laser wavelength and output power of the SBS micro laser are less than 0.002 nm and 0.007 mW, respectively, within the observation time of 1 hour. Figure 5(d) shows the measured linewidth within 1 hour, with fluctuation of less than 45 Hz. It is worth mentioning that the whole experiment is carried out under conditions of room temperature and normal pressure, without any thermostatic or vibration isolation. The experimental results show that the Brillouin microcavity laser, after all-optically locking the pump laser to the microresonator by self-injection, can guarantee remarkable long-term stability with mode-hop-free operation for at least 1 hour.

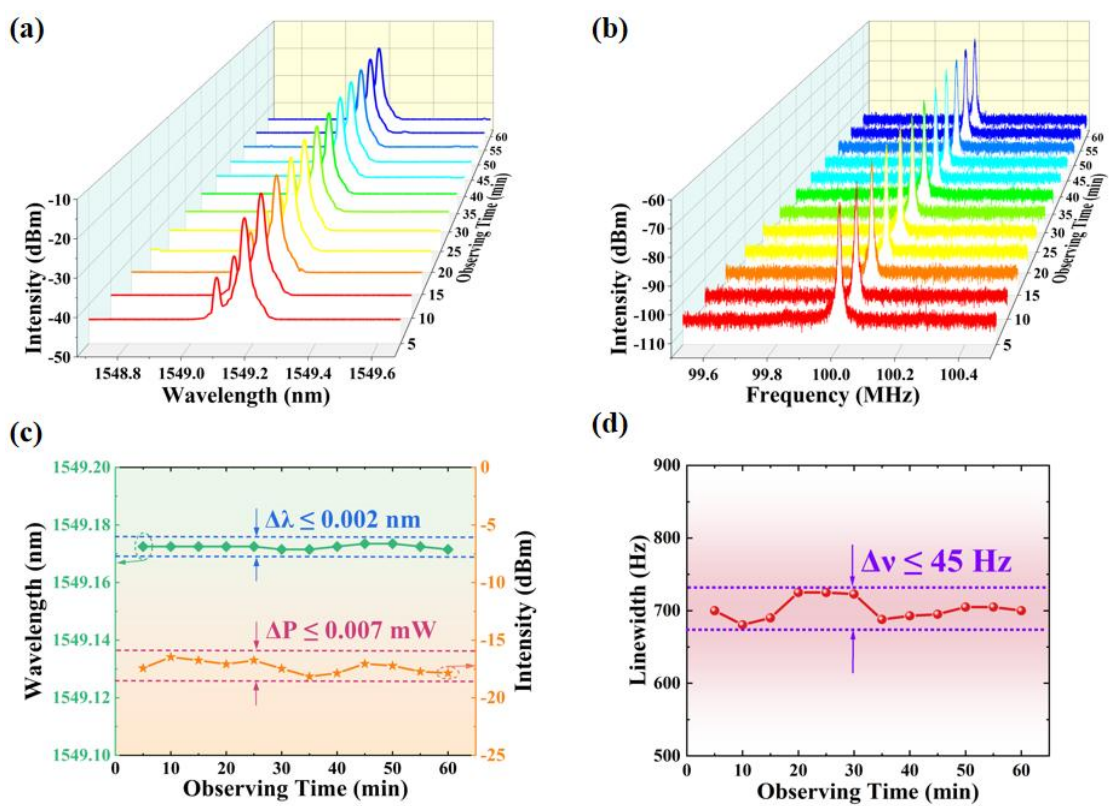

Fig. 5. Long-term stability of the Brillouin microlaser. (a) Optical spectrum measurement within 1 h. (b) Linewidth measurement within 1 h. (c) Wavelength and intensity stability, and (d) linewidth stability within 1 h.

4. Conclusion

In summary, the long-term stability of the microcavity Brillouin laser has been greatly improved by locking the pump laser to the microcavity with self injection from Rayleigh scattering. The pump-locked state of the constructed Brillouin microlaser can be maintained for at least 1 hour under normal laboratory conditions. At the same time, The pump and microcavity Brillouin lasers with ultra-low frequency noise floors of 7 Hz$^2$/Hz and 35 Hz$^2$/Hz have been realized, respectively. In addition, the wavelength of the Brillouin microlaser can be precisely tuned by adjusting the wavelength and power of the pump laser. To the best of our knowledge, this work is the first demonstration of an all-optically pump-locked microcavity Brillouin laser. The proposed pump-locking method does not require expensive narrow-linewidth tunable lasers and greatly improves the stability of optically pumped microcavity lasers, reducing the complexity and cost of practical microlaser systems. The constructed pump-locked microcavity Brillouin laser provides a stable and reliable low-cost experimental platform for on-chip integration of ultra-narrow linewidth lasers, precision laser sensors, microwave photonics signal synthesizers and optomechanical systems.

Funding

This work was supported in part by the National Natural Science Foundation of China (NSFC) under Grants U23A20378, 61935007 and 62075020, in part by the Chongqing Natural Science Foundation of Innovative Research Groups under Grant cstc2020jcyj-cxttX0005, and in part by the National Science Fund for Distinguished Young Scholars under Grant 61825501.